\documentclass[a4paper,fleqn,usenatbib]{mnras}

\usepackage[T1]{fontenc}
\usepackage{ae,aecompl}

\usepackage{graphicx}	
\usepackage{amsmath}	
\usepackage{amssymb}	
\usepackage{multirow}

\newcommand{\HI}{H\,{\sc i}}
\newcommand{\Ht}{H$_2$}
\newcommand{\kms}{km\,s$^{-1}$}
\newcommand{\msun}{M$_{\odot}$}

\title[HCG~44]{\HI\ in Group Interactions: HCG~44}

\author[K. M. Hess et al.]{
Kelley M.~Hess$^{1,2}$,\thanks{E-mail: hess@astro.rug.nl}
M.~E.~Cluver$^{3}$, 
Sahba Yahya$^{3}$, 
Lukas Leisman$^{4}$, 
Paolo Serra$^{5}$, 
\newauthor
Danielle M.~Lucero$^{1}$, 
Sean S.~Passmoor$^{6}$, and
Claude Carignan$^{7,8}$
\\
$^{1}$Kapteyn Astronomical Institute, University of Groningen, Landleven 12, 9747 AD, Groningen, The Netherlands\\
$^{2}$ASTRON, the Netherlands Institute for Radio Astronomy, Postbus 2, 7990 AA, Dwingeloo, The Netherlands\\
$^{3}$Physics Department, University of the Western Cape, Cape Town 7535, South Africa \\
$^{4}$Cornell Center for Astrophysics and Planetary Science, Space Sciences Building, Cornell University, Ithaca, NY 14853, USA \\
$^{5}$CSIRO Astronomy and Space Science, Australia Telescope National Facility, PO Box 76, Epping, NSW 1710, Australia \\
$^{6}$Square Kilometre Array South Africa, The Park, Park Road, Pinelands 7405, South Africa \\
$^{7}$Department of Astronomy, University of Cape Town, Private Bag X3, Rondebosch 7701, South Africa \\
$^{8}$Laboratoire de Physique et Chimie de l'Environnement (LPCE), Observatoire d'Astrophysique de l'Universit\'e de Ouagadougou, BP 7021, \\
Ouagadougou 03, Burkina Faso
}

\date{Accepted 2016 September 13. Received 2016 September 12; in original form 2016 May 26}

\pubyear{2016}

\begin{document}
\label{firstpage}
\pagerange{\pageref{firstpage}--\pageref{lastpage}}
\maketitle

\begin{abstract}
Extending deep observations of the neutral atomic hydrogen (\HI) to the environment around galaxy groups can reveal a complex history of group interactions which is invisible to studies that focus on the stellar component.  Hickson Compact Group 44 (HCG~44) is a nearby example and we have combined \HI\ data from the Karoo Array Telescope, Westerbork Synthesis Radio Telescope, and Arecibo Legacy Fast ALFA survey, in order to achieve high column density sensitivity ($N_{HI}<2\times10^{18}$ cm$^{-2}$) to the neutral gas over a large field-of-view beyond the compact group itself.  
We find the giant \HI\ tail north of HCG~44 contains $1.1\times10^9$~\msun\ of gas and extends 450 kpc from the compact group: twice as much mass and 33\% further than previously detected.  However, the additional gas is still unable to account for the known \HI\ deficiency of HCG~44.  The tail likely formed through a strong tidal interaction and \HI\ clouds in the tail have survived for 1 Gyr or more after being stripped.  This has important implications for understanding the survival of neutral clouds in the intragroup and circumgroup medium, and we discuss their survival in the context of simulations of cold gas in hot halos.
HCG~44 is one of a growing number of galaxy groups found to have more extended \HI\ in the intragroup and circumgroup medium than previously measured.  Our results provide constraints for simulations on the properties of galaxy group halos, and reveal a glimpse of what will be seen by future powerful \HI\ telescopes and surveys.\end{abstract}




\section{Introduction}

Large area redshift surveys such as SDSS \citep{Ab09} and GAMA \citep{Dr09} have established a new paradigm regarding the clustering of galaxies \citep{Ber06,Rob11}. The evolutionary effects imposed by this range of environments are not fully understood, largely due to the complexity of N-body interactions on star formation and gas dynamics. However, the relatively unexplored regime of neutral \HI\ in the context of galaxy groups has the potential to revolutionise our understanding of the mechanisms influencing how galaxies mature. 

Targeted observations of the \HI\ content in nearby groups have often revealed the presence of intragroup material, for example, with the Very Large Array \citep[VLA e.g.][]{van88,Yun94}, the Parkes radio telescope \citep[e.g.][]{Kil06}, the Australia Telescope Compact Array \citep[e.g.][]{Bar01, Kor04}, and Nan\c{c}ay \citep{vDr01}. However, systematic surveys of galaxy groups are rare due to the observational time necessary to achieve the required sensitivities, even at relatively modest redshifts, over large fields of view. The blind, wide-area Arecibo Legacy Fast ALFA (ALFALFA) survey \citep{Gio05,Hay11,AL12}, combined with the Sloan Digital Sky Survey (SDSS) DR7 group catalogue \citep{Ber06,Ber09}, enabled \citet{HW13} to explore the \HI\ content of group galaxies as a function of optical group membership.  They found evidence for gas pre-processing correlated with group assembly. Meanwhile, the Blind Ultra Deep HI Survey (BUDHIES) on the Westerbork Synthesis Radio Telescope (WSRT), a targeted survey at $z\approx0.2$, combines accurate \HI\ content with morphology in environments ranging from voids to cluster centres to show evidence for both ram pressure and pre-processing as mechanisms for gas removal \citep{Jaf13,Jaf15}.

The role of compact groups, in particular, has long been advocated for understanding galaxy evolution processes: their high galaxy densities combined with relatively shallow potential wells prolongs strong gravitational interactions. \citet{Hick82} defined compact groups as having 4 or more members within a 3 magnitude range ($\delta$mB) that form a physically isolated system. The original Hickson catalogue has been modified based on radial velocity information with 92 groups having at least three accordant members \citep{Hick92}.  Although other compact group catalogues exist, it remains the most intensively studied sample.

The \HI\ deficiency in compact group galaxies \citep{Will87,Huch97} was the focus of a single-dish study of 72 HCGs by \citet{VM01}.  In combination with a subset resolved by the VLA, the authors proposed an evolutionary sequence where galaxies become increasingly \HI\ deficient due to multiple tidal interactions.  The study of compact group evolution was expanded by Green Bank Telescope (GBT) observations from \citet{Bor10} which suggested the presence of copious diffuse, low column-density neutral hydrogen gas between the galaxies.  However, even this emission is not always able to account for the \HI\ deficiency of compact groups.   Further studies suggest that the most evolved groups, which are also dominated by elliptical galaxies, begin to exhibit signs of hot intragroup gas due to gravitational heating \citep{Pon96,Os04}.  However, strong \HI\ deficiency does not guarantee the presence of intragroup X-ray emitting gas, complicating the interpretation of what mechanisms are responsible for removing the cold gas \citep{Ras08,Des12}. 

Additional evidence of evolution within compact groups has come from photometry from the {\it Spitzer} Space Telescope: bimodality in IRAC color-color space, appearing to correlate with \HI\ depletion, suggests accelerated evolution from dusty, star-forming galaxies, to gas-poor systems \citep{Joh07, Walk10}. Detailed studies by \citet{VM98,Tzan10,Walk12} further support a picture of accelerated evolution. Evidence of shock-excited \Ht\ in several compact group galaxies led \citet{Clu13} to suggest a connection between intragroup tidal debris and rapid quenching in these groups.

HCG 44 is a relatively nearby compact group with redshift $z \sim 0.0046$ and an angular diameter of approximately $16$\arcmin.  It contains four massive galaxies: three star-forming spirals (NGC~3185, NGC~3187, NGC~3190), and one elliptical which is undetected in \HI\ (NGC~3193); and the \HI\ cloud, $C_{\text{S}}$, which lacks an apparent optical counterpart \citep{Ser13}.  The galaxies are highly \HI-deficient \citep{VM01}, but X-ray emission is confined to individual galaxies with no detectable diffuse X-ray emission associated with the intragroup medium \citep{Ras08}, leaving the missing \HI\ unaccounted for.  Further, the lack of diffuse hot gas brings into question the importance of ram pressure stripping as a mechanism for removing gas from the large galaxies in this system.  \citet{Clu13} found excited \Ht\ in NGC~3190, perhaps shocked from impact with the intragroup medium, but GBT observations are unable to resolve individual galaxies, and high resolution interferometry observations have not yet detected intragroup \HI\ within the core of the compact group.  In fact, the evolutionary stage of the compact group has been debated as the \HI\ content suggests it is more advanced, while the lack of intragroup optical light from stars, and intragroup X-rays suggest it is younger \citep{Agu06,Os04}.  

As part of the ATLAS$^{\rm 3D}$ survey, \citet{Ser12} surveyed the neutral gas content of HCG 44 and found evidence for an extended \HI\ distribution well beyond the compact group. Deeper WSRT data, combined with reprocessed HIPASS data, detected the presence of a $\sim300$\,kpc long tail to the north of HCG~44 containing $5\times10^{8}M_{\sun}$ of \HI\ \citep{Ser13}.  The authors concluded that this tail is most likely the result of tidal interactions and proposed possible scenarios.  First, the tail is the result of a high speed interaction between NGC 3162 and the compact group which occurred up to 3 Gyr in the past.  NGC~3162 lies $\sim$1.4 deg (620 kpc at the assumed distance) northwest of HCG~44.  Second, the tail is a result of tidal stripping due to NGC 3187 interacting with the gravitational potential of the compact group, which would explain its very disturbed optical morphology.  However, they noted that deeper \HI\ data over a larger field would benefit detailed modeling of this system.

In this paper we investigate the extended region around HCG~44 and the northern tail by conducting a deep \HI\ mosaic with the Karoo Array Telescope (KAT-7) which, due to its sensitivity and resolution, is ideal for detecting low column density \HI\ gas \citep{Car13}.  We combine these data with the deep WSRT pointing from \citet{Ser13}, and an \HI\ map spanning the same sky coverage from the ALFALFA survey to improve our overall picture of the \HI\ distribution in the compact group, and nearby surroundings.

The paper is organized as follows. In Section \ref{data} we describe the KAT-7 observations, and summarize the additional data.  In Section \ref{combo} we describe our technique for combining interferometric images from KAT-7 and WSRT, and the single dish map from ALFALFA.  In Section \ref{results} we compare the measurements from KAT-7 and ALFALFA and present our results from the combined data cube.  In Section \ref{discussion} we discuss the possible origins and the fate of the extended \HI\ tail.  We conclude in Section \ref{conclusions}.

\begin{figure*}
\includegraphics[scale=0.89,clip,trim=20 70 0 80]{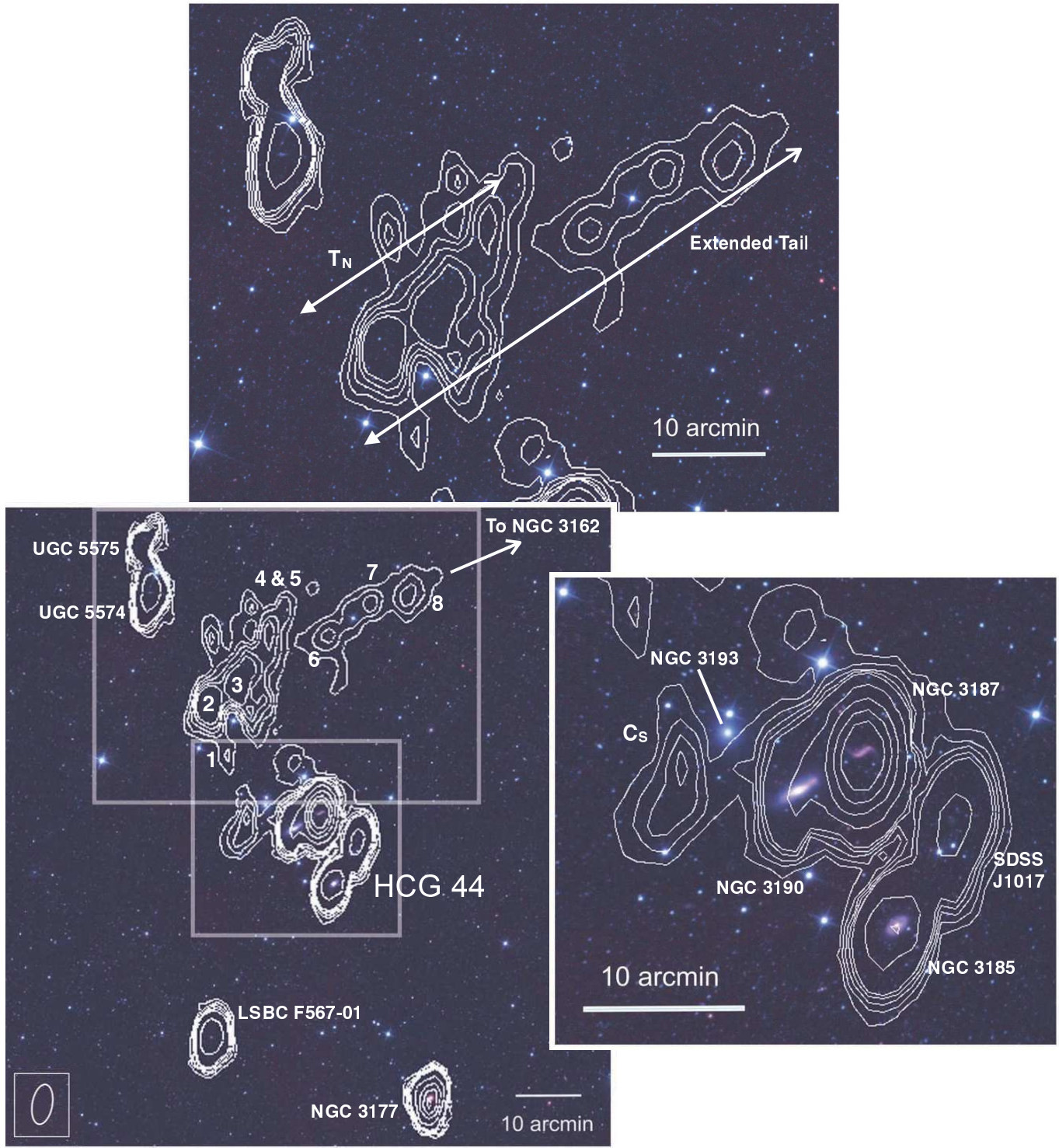}
\caption{A ``drizzled'' \textit{Wide-field Infrared Survey Explorer} (WISE) four-color image of HCG~44 and the extended \HI\ tail: 3.4 $\mu$m (W1) is blue, 4.6 $\mu$m (W2) is green, 12 $\mu$m (W3) is orange, and 22 $\mu$m (W4) is red.  White contours correspond to \HI\ column densities of 2, 4, 6, 8, 10, 25, 50, 75, $100\times10^{18}$ cm$^{-2}$ from the KAT-7 observations, and we have labeled the major \HI\ features.  \HI\ clouds 4 and 5 in $T_{\text{N}}$ are not well resolved in KAT-7, but are resolved in WSRT data (see Figure \ref{mom0}).  Star forming galaxies appear red.  Foreground Milky Way stars and old stellar populations appear blue.  There is no evidence for a diffuse stellar counterpart to the \HI\ in WISE imaging outside of the known galaxies.}
\label{wise}
\end{figure*}

\section{Data} \label{data}

\subsection{KAT-7 observations, and data reduction}

HCG~44 and the \HI\ tail ($T_{\text{N}}$; \citealt{Ser13}) were observed with the KAT-7 telescope \citep{Foley} in two pointings: one centered on the group itself ($\alpha=10^{\text{h}}17^{\text{m}}46.03^{\text{s}}\,+21^{\circ}48^{\prime}30.8^{\prime\prime}$; hereafter Field 1), and one centered on the northern tail discovered by \citet{Ser13} ($\alpha=10^{\text{h}}17^{\text{m}}54.312^{\text{s}}\,+22^{\circ}20^{\prime}37.1^{\prime\prime}$; hereafter Field 2). Field 1 was observed with all seven antennas in 11 sessions between 2015 Jan 9 -- 2015 May 20 for a total of 30 hours on source.  Field 2 was later observed with six available antennas in 8 sessions between 2015 July 3 -- 2015 July 13 for a total of 36.25 hours on source.  Field 1 was observed during the night, whereas Field 2 had primarily day time observations and was affected by solar interference which resulted in more intensive flagging during data reduction.

The KAT-7 data were reduced in CASA 4.4.0 \citep{McMul07} using standard procedures.  As a first step, we flagged all $u=0$ data to remove low level radio frequency interference (RFI) that is correlated when the fringe rate is equal to zero (see \citealt{Hess15} for a complete explanation).  
After flux and gain calibration, we performed continuum subtraction using a first order fit in the task UVCONTSUB and imaged each observation individually for quality assurance.  Field 1 and Field 2 achieved an rms noise of 2.7 and 2.9 mJy~beam$^{-1}$, respectively, over 7.7~\kms\ channels, despite the differing total observing times between the fields.  We performed self-calibration on the continuum data of the target field and applied it to the spectral line data.  This generally produced a factor of 1.2 improvement in the rms noise, and up to a factor of 1.5 in individual data sets.  The final mosaic improved to 1.7 mJy~beam$^{-1}$, with a $358\times187$ arcsec beam, thanks to self calibration and overlap between the fields.  

We converted this to column density using Equation 1 from \citealt{Hib01}:
\begin{equation}
N_{HI}=\frac{1.104\times10^{21} \textrm{cm}^{-2}}{\theta_a\times\theta_b}\frac{\int S \Delta v}{\textrm{mJy~beam}^{-1} \textrm{km~s}^{-1}},
\end{equation}
where $S$ is in mJy, $\theta_a$ and $\theta_b$ are the major and minor axes of the beam in arcseconds, and $\Delta v$ is the channel width in \kms.  This corresponds to a 1 $\sigma$ \HI\ column density of $2.2\times10^{17}$ cm$^{-2}$ channel$^{-1}$.  Figure \ref{wise} shows the \HI\ contours of the KAT-7 mosaic overlaid on a four-color image from the \textit{Wide-field Infrared Survey Explorer} (WISE).  We describe the method for generating these contours in Section \ref{results}.

\subsection{WSRT data}

\cite{Ser13} observed HCG~44 with the WSRT for $6\times12$ hours at a single pointing centered between the group and the \HI\ tail ($\alpha=10^{\text{h}}18^{\text{m}}45.19^{\text{s}}\,+21^{\circ}59^{\prime}30.5^{\prime\prime}$.  The observations achieved an rms of 0.23 mJy~beam$^{-1}$ with a beam of $53.0\times32.7$ arcsec over 8.2~\kms\ channels.

\subsection{ALFALFA data}

HCG~44 also falls in the portion of sky observed by the ALFALFA survey.  A full description of the survey strategy can be found in \citet{Gio05} and \citet{Hay11} but, in short, ALFALFA is a two-pass, fixed-azimuth, drift scan survey covering the velocity range from $-300$ to $+18,000$~\kms.  We obtained a data cube spanning $10^{\text{h}}14^{\text{m}}26^{\text{s}}\,+21^{\circ}03^{\prime}22^{\prime\prime}$ to $10^{\text{h}}21^{\text{m}}11^{\text{s}}\,+23^{\circ}04^{\prime}08^{\prime\prime}$ and $700-2010$~\kms.  The cube has an overall rms of 2.67 mJy beam$^{-1}$ with a $228\times210$ arcsec beam and 5.15~\kms\ channels.  We note that the data cube was affected by low level RFI in the declination strip that passed through the compact group itself \citep{Lei16}, but we are able to take this into consideration with the weights map that is simultaneously created in the ALFALFA data reduction process.

\section{Data combination} \label{combo}

Traditional methods of combining radio data between different arrays, or between arrays (interferometers) and single-dish measurements, may employ methods which operate in either the ($u$,$v$)-plane or in the image plane (\citealt{Stan02} and references therein).  The contribution of each measurement is generally weighted by the inverse square of the rms, $1/\sigma_{rms}^2$.  In the case of HCG~44, the WSRT observations are an order of magnitude more sensitive than the KAT-7 data in Jy beam$^{-1}$, making an attempt at traditional combination relatively fruitless.

However, recognizing that the WSRT beam is more than an order of magnitude smaller in area than the KAT-7 beam, \citet{Sor15} demonstrate an innovative and powerful method of combining KAT-7 and WSRT mosaics to reveal the presence of a low surface brightness tail in Virgo galaxy, NGC~4424.  Their technique realizes the strengths of both telescopes by converting the cubes to column density units which removes the dependence of the measured rms on the beam size, and allows the image cubes to be combined with more equal weights.  As a result, the final image of NGC~4424 has the advantage of sensitivity to large-scale structure from the short baselines of KAT-7, and high resolution from WSRT.  The full procedure is described in \citet{Sor15} and we highlight the details relevant to the successful combination of our data below.

We employed the same technique to combine the interferometry data with the ALFALFA maps.  The most common method of combining interferometry and single-dish observations is to ``feather'' the data sets.  This refers to performing a Fourier Transform (FT) on the individual data to the $(u,v)$-plane, scaling and combining them, then performing an inverse FT back to the image plane \citep{Stan02}.  This is typically done because the single-dish and interferometers measure different spatial scales, and the single-dish data is being combined to get the zero-spacing measurement.  However, the short baselines of KAT-7 (26--185 m) measure effectively the same spatial scales as Arecibo (illuminated diameter of 208 m), and the ALFALFA drift scans have been synthesized to data cubes from which to extract \HI\ detections.  Thus, we use the ALFALFA beam information to convert the ALFALFA cube to column density units, and combine with the KAT-7 and WSRT data in the image plane.  Formally, as long as the data is weighted appropriately, the methods produce comparable quality results \citep{Stan02}.

\begin{figure}
\includegraphics[scale=0.46,clip,trim=33 85 50 40]{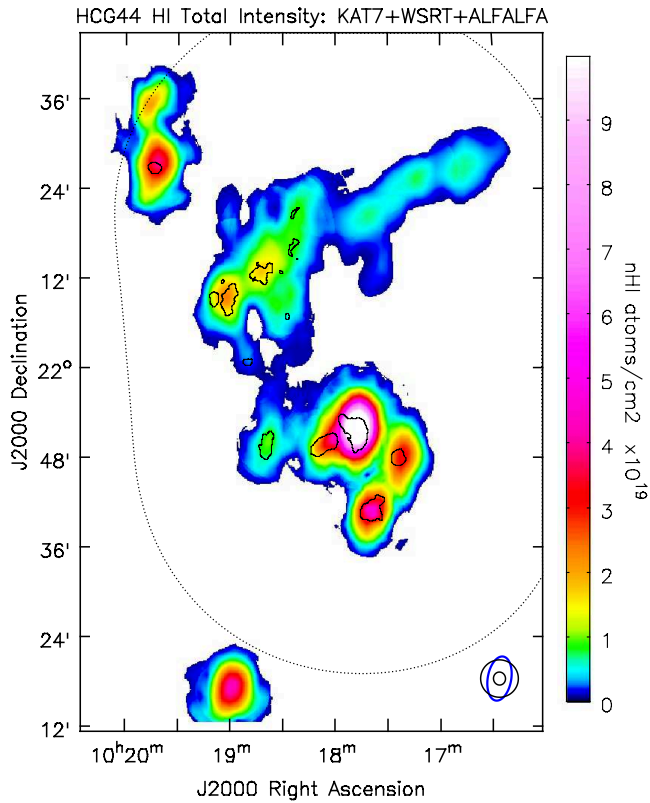}
\caption{The \HI\ total intensity map of the combined KAT-7, ALFALFA, and WSRT observations (Equation \ref{finalcube}). The black contour corresponds to the $1\times10^{19}$ cm$^{-2}$ column density contours of the original WSRT observation \citep{Ser13}. The dotted line is the half power of the KAT-7 mosaic.  The beams from each telescope are shown in the lower right hand corner, in order of decreasing size: KAT-7 (elongated), ALFALFA (round), smoothed WSRT.  The WSRT and ALFLFA data cubes we acquired do not extend as far south as the KAT-7 mosaic, so NGC~3177 does not appear in this image.}
\label{mom0}
\end{figure}

\subsection{Constructing the final cube} 

To optimize the final weighting scheme versus desired resolution, the WSRT cube from \citet{Ser13} was smoothed with a $90\times90$ arcsec Gaussian kernel resulting in an effective restoring beam of $104\times95.7$ arcsec and rms noise of 0.39 mJy beam$^{-1}$.  As further preparation, the KAT-7, WSRT, and ALFALFA cubes were smoothed in velocity using a boxcar kernel of 2, 2, and 3 channels, respectively, and decimated in order to achieve the lowest multiple common velocity resolution in independent channels.
The KAT-7 and ALFALFA cubes were then regridded to match the velocity and spatial axes of the WSRT cube, so that the resulting cubes have the same pixel size and dimensions ($8^{\prime\prime}\times8^{\prime\prime}\times16$~\kms).  All cubes were converted to column density per unit velocity by the following equation:
\begin{equation}
N_{HI}=1.1\times10^{21}\left (\frac{I^{i,j}_{HI}}{b_{max}b_{min}}\right )
\end{equation}
where $I^{i,j}_{HI}$ is the \HI\ intensity of pixel ($i$,$j$) in mJy~beam$^{-1}$~\kms, and $b_{max}$ and $b_{min}$ are the major and minor axes of the restoring beam in arcseconds.  The resulting KAT-7, WSRT, and ALFALFA data cubes have rms noise of $1.48\times10^{16}$, $4.00\times10^{16}$, and $3.39\times10^{16}$ cm$^{-2}$(\kms)$^{-1}$,  respectively.

Before linear combination, the KAT-7 and WSRT mosaics were primary beam corrected.  CASA automatically generates a primary beam pattern when the visibilities are imaged, which we used to correct the KAT-7 data.  With only the final WSRT image cube, we assumed the primary beam correction to be a Gaussian profile with 35 arcmin width. The WSRT data was imaged well beyond the primary beam of the telescope, and if combined linearly with the KAT-7 and ALFALFA data, the final image would be WSRT noise dominated at large distances from the WSRT pointing center.  Thus, we only combine with WSRT where the primary beam sensitivity is at least 30\% of the peak.

We experimented with combining the image cubes with increasingly complex set of weights to take into account different considerations.  Most simply, the KAT-7 and WSRT primary beam corrected cubes, and ALFALFA cube were weighted by the square of the rms and combined such that:
\begin{equation}
I_\text{KWA} (\text{PB}_\text{W} > 30\%) = \frac{I_\text{KAT7}+0.14\,I_\text{WSRT}+0.19\,I_\text{ALFA}}{1.33}
\end{equation}
\begin{equation}
I_\text{KWA} (\text{PB}_\text{W} < 30\%) = \frac{I_\text{KAT7}+0.19\,I_\text{ALFA}}{1.19}
\end{equation}

The resulting cube is dominated by the KAT-7 observations throughout.  

Despite the relatively large field of view of KAT-7 (approximately 1 degree at the half power beam width), it also falls off in sensitivity with distance from the mosaic pointing centers, while the ALFALFA sensitivity is fairly flat across the field.  Thus, in our final image combination, we take into account not only the relative rms noise at the most sensitive part of the images, but the change in sensitivity across the field of view due to the primary beam \citep{Condon98}.  We also consider the ALFALFA weights map such that:
\begin{equation}
I_\text{KWA} = \frac{PB^2_\text{K}\,I_\text{K}+0.14\,PB^2_\text{W}\,I_\text{W}+0.19\,w_\text{A}\,I_\text{A}}{PB^2_\text{K}+0.14\,PB^2_\text{W}+0.19\,w_\text{A}}.
\label{finalcube}
\end{equation}

This cube is dominated by the KAT-7 observations at the center and the ALFALFA observations at the outskirts.  This also means that the effective synthesized beam changes across the field.  However, because the final beam size of KAT-7 and ALFALFA are similar, we do not consider it to be a significant complication to the interpretation of the final data cube.  See the beams in Figure \ref{mom0}, for example.  The peak rms sensitivity of this final cube improves to $1.35\times10^{16}$ cm$^{-2}$(\kms)$^{-1}$ at a velocity resolution of 16.4~\kms.  The mean rms across the image is $2.43\times10^{16}$ cm$^{-2}$(\kms)$^{-1}$.

Perhaps the greatest benefit of this combination is that the KAT-7 and ALFALFA cubes individually contain residual artifacts whose patterns are characteristic to the telescope, but by combining them one is better able to differentiate between noise and diffuse low-level emission.  Throughout the rest of the paper, the \HI\ data presented are measured or calculated from the combined KAT-7, WSRT, ALFALFA (KWA) image cube, unless otherwise specified (Equation \ref{finalcube}).

\begin{table*}
	\caption{Properties of HI Detections}
	\begin{tabular}{lccccccccc}
	\hline
	         & Opt.~Pos.         & Velocity & Dist. & GBT$^1$ & KAT-7     & KWA              & ALFALFA$^2$        & VLA-D$^3$  & WSRT$^4$    \\
	Name     &  J2000            &  \kms    & Mpc   & \multicolumn{2}{c}{$\log(M_{HI}/M_{\odot})$}         & \multicolumn{2}{c}{$\log(M_{HI}/M_{\odot})$} & \multicolumn{2}{c}{$\log(M_{HI}/M_{\odot})$} \\
	\hline
	NGC 3185 & 10:17:38.5 +21:41:18 & 1210     & 25 & --   & 8.58         & 8.58             & 8.72\phantom{$^*$} & 8.68$^{\dagger}$ & 8.49\phantom{$^*$} \\
	NGC 3187$^*$ & 10:17:47.8 +21:52:24 & 1575 & 25 & --   & \multirow{2}{*}{9.23} & \multirow{2}{*}{9.25} & 9.22\phantom{$^*$} & 9.19$^{\dagger}$ & 9.08\phantom{$^*$} \\
	NGC 3190$^*$ & 10:18:05.6 +21:49:56 & 1300 & 25 & --   &              &                  & 8.82\phantom{$^*$} & 8.88$^{\dagger}$ & 8.76\phantom{$^*$} \\
	$C_{\text{S}}$ &                & 1405     & 25 & --   & 8.05         & 8.04             & --\phantom{$^*$}   & --               & 8.08\phantom{$^*$} \\
	HCG 44   &           & --       & 25 & 9.46$^{\dagger}$ & 9.34        & 9.35             & 9.45\phantom{$^*$} & 9.44$^{\dagger}$ & 9.34\phantom{$^*$} \\
	$T_{\text{N}}$ &   & \multirow{2}{*}{1260-1365} & 25 & --   & 8.89         & 8.88             & --\phantom{$^*$}   & --               & 8.61\phantom{$^*$} \\
	Extended Tail &                 &          & 25 & --   & 9.03         & 9.02             & 9.05\phantom{$^*$} & --               & --\phantom{$^*$}   \\
        \hline       
        NGC 3177   & 10:16:34.1 +21:07:23 & 1295   & 25 & --   & 8.90         & --               & 8.74\phantom{$^*$}   & --               & --\phantom{$^*$}   \\
	SDSS J1017 & 10:17:23.3 +21:47:58 & 1930   & 29 & --   & 8.88         & 8.88             & 8.80$^{\dagger}$   & --               & 8.30\phantom{$^*$} \\
	LSBC F567-01 & 10:19:01.5 +21:17:01 & 1075 & 18 & --   & 8.37         & 8.37             & 8.01$^{\dagger}$   & --               & --\phantom{$^*$}   \\
	UGC 5574  & 10:19:42.9 +22:27:09 & 1460    & 25 & --   & 8.61         & 8.62             & 8.53\phantom{$^*$} & --               & 8.12$^{\dagger}$   \\
	UGC 5575  & 10:19:46.8 +22:35:41 & 1460    & 25 & --   & 8.19         & 8.21             & 8.28\phantom{$^*$} & --               & --\phantom{$^*$}   \\
	\hline
	\end{tabular} \\
	$^*$NGC~3187 and NGC~3190 are blended in the KAT-7 data.  $^{\dagger}$\HI\ masses from the literature are corrected to our assumed distance.  Where the total \HI\ mass is not reported in the literature, we add the masses from individually reported group members. The telescope columns are listed in order of decreasing beam size.  $^1$15 arcmin beam; \citet{Bor15}. $^2$$3.5\times3.8$ arcmin beam; \citet{Lei16}. $^3$$60\times61$ arcsec beam; we have subtracted HCG~44e from the total \HI\ mass reported for HCG~44 because it is background galaxy at 4032\kms; \citet{VM01}. $^7$$53\times33$ arcsec beam; \citet{Ser13}.
	\label{sources}
\end{table*}

\section{Results} \label{results}

In Figure \ref{mom0}, we present the \HI\ total intensity map derived from the combined KWA data cube.  The map was made by smoothing the cube with a $5\times5$ arcmin beam, and masking regions where the emission was greater than 1.5$\sigma$ in the original cube.  The mask was inspected and modified to keep regions with emission spatially contiguous over at least 2 channels, and remove peaks suspected to be residual RFI.  We took particular care to include even potential emission in the region between HCG~44 and the northern \HI\ tail (see Section \ref{origins} for further explanation).  The mask was similarly inspected and applied to the KAT-7 data and was used to generate the contours overlaid in Figure \ref{wise}.

The addition of the KAT-7 and ALFALFA observations shows that the \HI\ tail extends further than previously detected with WSRT.  Based on low signal-to-noise HIPASS images, \citet{Ser13} found that the tail stretched 300 kpc in projection.  At the assumed distance of 25 Mpc (see Section \ref{dists}), we find the extended tail spans nearly 450 kpc (62 arcmin) in projection and is arguably resolved into seven to eight clumps.  The extent is measured from the southern extension (cloud 1), through the brightest part of the tail, to the northwestern cloud 8.  The physical length of the tail may be even longer than what is seen in projection.  Our images reveal the breadth, mass, column density, and velocity dispersion of the tail that was unresolved by HIPASS.  Throughout the rest of this work, we will refer to the tail resolved by the WSRT maps as $T_{\text{N}}$, and the full extent revealed by the KWA combined data, which includes $T_{\text{N}}$, as the ``extended tail.''

In addition to HCG~44, we detect the suspected background galaxy, SDSS~1017723.29+214757.9 (hereafter SDSS~1017; \citealt{Ser13}), and likely foreground galaxy, LSBC~F567-01.  We note that the elliptical galaxy, NGC~3193, remains undetected in \HI.  \citet{Agu06} propose that despite, its velocity (1381~\kms), the absence of planetary nebulae in narrow band observations suggest it lies behind the compact group.  Finally, UGC~5574 and UGC~5575 are also detected and, while not formally part of HCG~44, are likely part of the larger associated group of galaxies \citep{Mak11}.

We present the properties of all the \HI\ detections in Table \ref{sources} with comparison to previous detections.  In order to convert from column density to solar masses we use 1~\msun~pc$^{-2} = 1.248\times10^{20}$ atoms cm$^{-2}$, integrate over the volume of the emission in the cube, and use the pixel scale at the luminosity distance of the object to convert to parsec$^2$.  For the literature values, we take the published integrated flux and convert to \HI\ mass using our assumed distances, and the standard equation:
\begin{equation}
M_{HI}/M_{\odot} = 2.36\times10^5 D_\text{Mpc}^2 \int S_{21}\,dv
\end{equation}
where $\int S_{21}\,dv$ is the intergrated flux in Jy~\kms, $D_\text{Mpc}$ is the distance in Mpc, and $M_\text{\HI}$ is the \HI\ mass computed in units of solar mass.  These values are given in Table \ref{sources}.

\subsection{A note on distances}
\label{dists}

The greatest source of error on the true \HI\ mass content is the uncertainty in distance to individual galaxies both in and around the compact group.  NED gives estimates corrected for Virgo infall of order $22-25\pm2$ Mpc, while distances assumed in the literature range from 18.4 to 25 Mpc \citep{VM01,Bor15,Will91,Ser13}.  The NGC 3190 group itself is part of a broader filamentary structure, so distance estimates from recessional velocities can be significantly distorted by infall onto the filament.  Following the convention of WSRT and ALFALFA studies, we assume a distance of 25 Mpc to HCG~44 member galaxies NGC~3185, NGC~3187, NGC~3190, to the cloud, $C_{\text{S}}$, and the extended tail.  We assume the same for NGC~3177, UGC~5574, and UGC~5575 which have similar recessional velocities and are at the same distance, within uncertainties, in velocity flow models of the region (see Table \ref{sources}).  
\citet{Lei16}) present a detailed discussion of distances to galaxies in the region.

\subsection{KAT-7 versus ALFALFA masses} \label{kva}

Table \ref{sources} provides a comparison of the \HI\ masses between the KAT-7 observations, the KWA results, and previous observations from a range of telescope facilities.  A cursory look reveals that the interpretation of fluxes from different facilities is not straight forward.  The amount of \HI\ detected by a given telescope is dependent on the size scales to which it is sensitive.  For example, if there is a significant amount of intragroup gas, we expect KAT-7 and ALFALFA to report more gas than WSRT, perhaps less gas than the GBT, and produce similar results to one another.  Errors in the absolute flux calibration would also produce different results between facilities.  Recalibrating data from five different telescopes to ensure consistency in the absolute flux calibration is outside the scope of this paper, however our observations give us the opportunity to comment on the relative flux calibration between KAT-7 and ALFALFA.

Among systematic sources of error, fitting a baseline to the data is particularly critical to account for both broadband continuum emission, and small scale residual bandpass fluctuations that may remain to separate the spectral line emission of individual galaxies \citep{vZee97}.  
For both KAT-7 and ALFALFA the baseline fitting takes place in a two step process.  
There is a first pass to remove broad continuum emission, and a second pass when the detections are measured which fits for local fluctuations of the baseline in emission free channels over the region in which the spectral line flux is extracted.  

\citet{Spr05} show that \HI\ flux densities measurements are typically not better than 15\%.  The mean difference between KAT-7 and ALFALFA \HI\ masses is 21\%.  However, there is not a simple universal offset.  Generally, the galaxies at the center of the KAT-7 mosaic (HCG~44 members and SDSS~J1017) are underestimated in the KAT-7 data compared to ALFALFA, and the galaxies on the outskirts of the mosaic (NGC~3177, LSBC~F567-01, and UGC~5574), where the sensitivity of the KAT-7 data are declining, are overestimated.  

\begin{figure}
\includegraphics[scale=0.48,clip,trim=30 10 5 25]{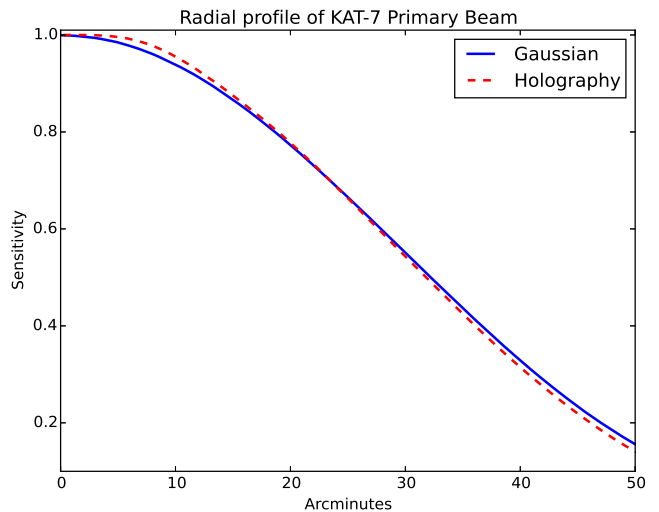}
\caption{A comparison of the assumed Gaussian primary beam used to correct the KAT-7 data (solid blue line) with engineering holography measurements of Antenna 1 Stokes I amplitude (dashed red line).  The holography data has been scaled radially to illustrate the approximate difference in primary beam correction which could account for the pattern of HI flux measurements.  The greatest difference for Antenna 1 is approximately 2\%, but may be greater for a full KAT-7 primary beam model.}
\label{beam}
\end{figure}

One possible explanation is that a better primary beam model is required for KAT-7 than the Gaussian assumed by the CASA software.  A full model for the KAT-7 primary beam is not currently available, but holography measurements of the individual dishes were taken as part of the commissioning process.  In Figure \ref{beam} we present the amplitude beam measurement of Antenna 1 Stokes I for KAT-7 compared to the Gaussian primary beam used to correct the data. The engineering holography measurement has been arbitrarily scaled in the radial direction for comparison.  We see that, indeed, it appears the Gaussian will under correct flux close to the center of a pointing, and over correct the flux on the outskirts.  For the Antenna 1 measurement, this difference is at the 2\% level, but it may be greater when the entire array is considered together.

\section{Discussion} \label{discussion}

A study of 22 HCGs, which included HCG~44, provided evidence of a diffuse neutral gas intragroup medium present in some compact groups \citep{Bor10}. However, based on follow-up mapping with the GBT, it was argued from the strong double-horned \HI\ profiles that the majority of \HI\ in HCG~44 is bound in stable disk-like structures \citep{Bor15}.  These observations covered the three main galaxies in HCG~44: NGC~3185, NGC~3187, NGC~3190; and the nearby $C_S$ cloud, and found a total \HI\ mass of $2.9\times10^9$~\msun\ but did not resolve the compact group.  We measure $2.3\times10^9$ \msun\ in HCG~44, which is consistent with the total \HI\ mass resolved by WSRT \citet{Ser13}.  While the GBT observations appear to detect 22\% more \HI\ in HCG~44 than either the KWA or WSRT observations, this seems to be within the variation measured for individual galaxies by different telescopes, even when those telescopes are sensitive to the same spatial scales.  The GBT and ALFALFA values are also consistent with the VLA-D observations, which show the \HI\ is confined to individual galaxies \citep{Will91}. Thus, we conclude that even with deeper \HI\ observations, there is no strong evidence for a diffuse, neutral IGM in HCG~44.

\subsection{\HI\ Deficiency in HCG~44}

HCG~44 is known to be highly \HI\ deficient, even when compared to other compact groups.  The expected \HI\ mass for the group is calculated by summing the predicted values from the individual galaxies based on their morphological type \citep{Hay84} and comparing them to the sum of the observed \HI\ masses: $Def_{HI} = \log[\Sigma M(HI)_{\text{pred}}]-\log[\Sigma M(HI)_{\text{obs}}]$.  The predicted value for HCG~44 is $\log[\Sigma M(HI)_{\text{pred}}]=10.2\pm0.2$ \citep{VM01}.  Our observed value is in line with previous measurements: $\log[\Sigma M(HI)_{\text{obs}}]=9.35\pm0.2$.

\citet{VM01} point out that the deficiency in HCG~44 is not due to a few highly anemic objects, but is a characteristic shared by all the members.  This argues that the group members have evolved together to account for the high \HI\ deficiency, and against a single interloper which has had its gas completely removed.  
The most probable effect produced by tidal interactions is believed to be disk stripping \citep{VM01}.  The role of ram-pressure stripping in compact groups has been shown to be small, only capable of removing small amounts of cold gas \citep{Ras08, Free10}, however, turbulent viscous stripping due to hydrodynamical interactions may also be an important mechanism for removing gas from compact group members \citep{Nul82,Quil00}.

\subsection{The Extended \HI\ tail} \label{hitail}

The advantage of low surface brightness sensitivity and wide field of view of our observations are apparent in observing the \HI\ tail, $T_{\text{N}}$, where we find $\sim$90\% more \HI\ than the WSRT observations alone.  However, the most striking result of our observations is detecting and resolving the gas beyond $T_{\text{N}}$.  The extended tail is resolved into 7-8 distinct knots, and the total \HI\ mass is of the order observed in a typical galaxy: $1.1\times10^9$~\msun.  Nonetheless, even including the extended tail, the total gas content of the HCG~44 system only adds up to $\log[\Sigma M(HI)_{\text{obs}}/M_{\odot}]=9.52\pm0.1$ and cannot account for the observed \HI\ depletion of the compact group.

\begin{figure}
\includegraphics[width=\columnwidth,clip,trim=20 200 50 100]{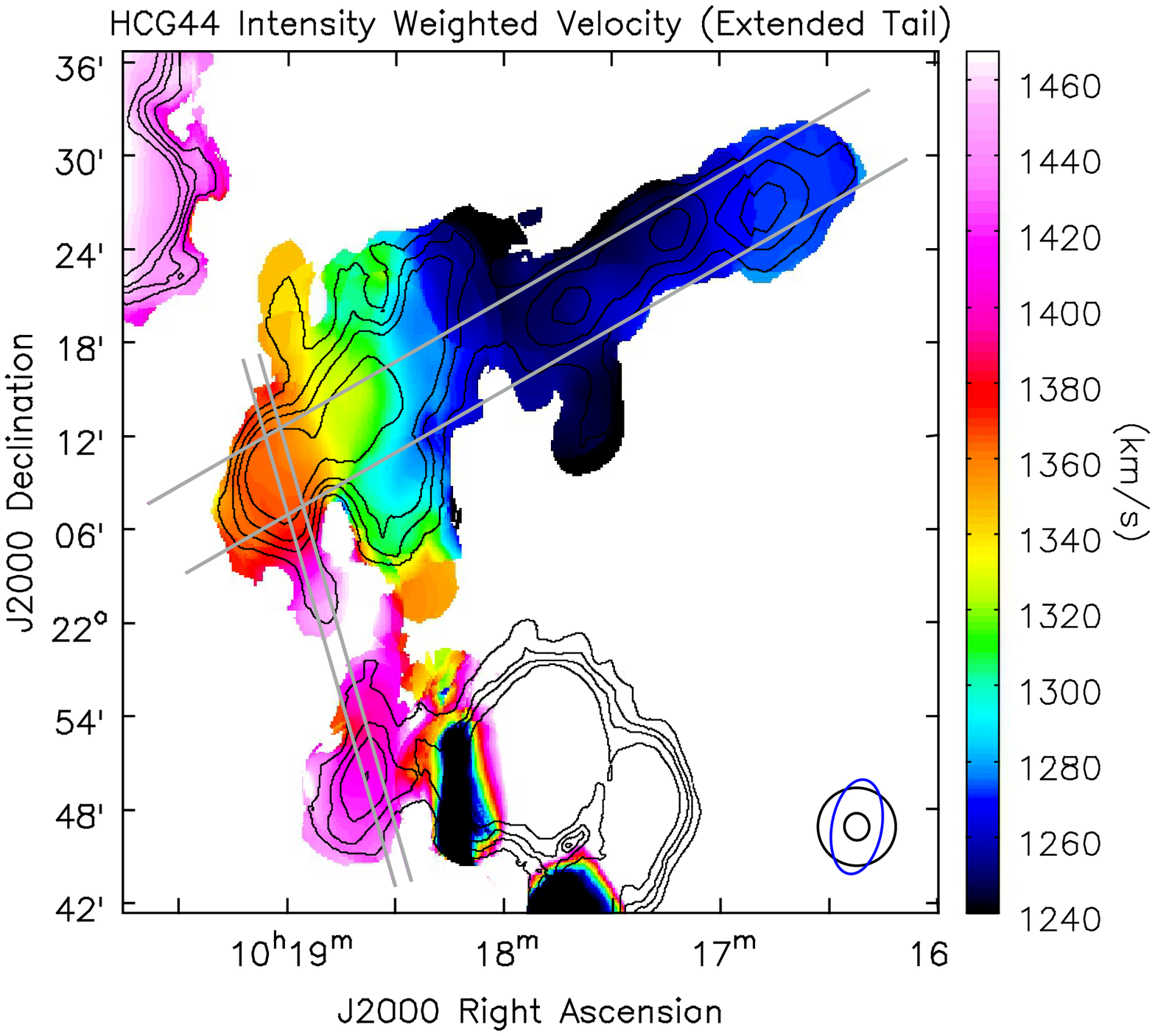}
\caption{The intensity-weighted velocity map of the \HI\ tail.  Although the tail has distinct peaks in intensity, it is continuous in velocity along the full length.  Contours are at 2, 4, 6, $10\times10^{18}$ cm$^{-2}$.  Parallel gray lines illustrate the orientation and width of the position-velocity slices in Figure \ref{pvmap} and \ref{pvmap2}. The beams for individual observations are shown in the lower right hand corner as in Figure \ref{mom0}.}
\label{tailvel}
\includegraphics[width=\columnwidth,clip,trim=40 330 32 150]{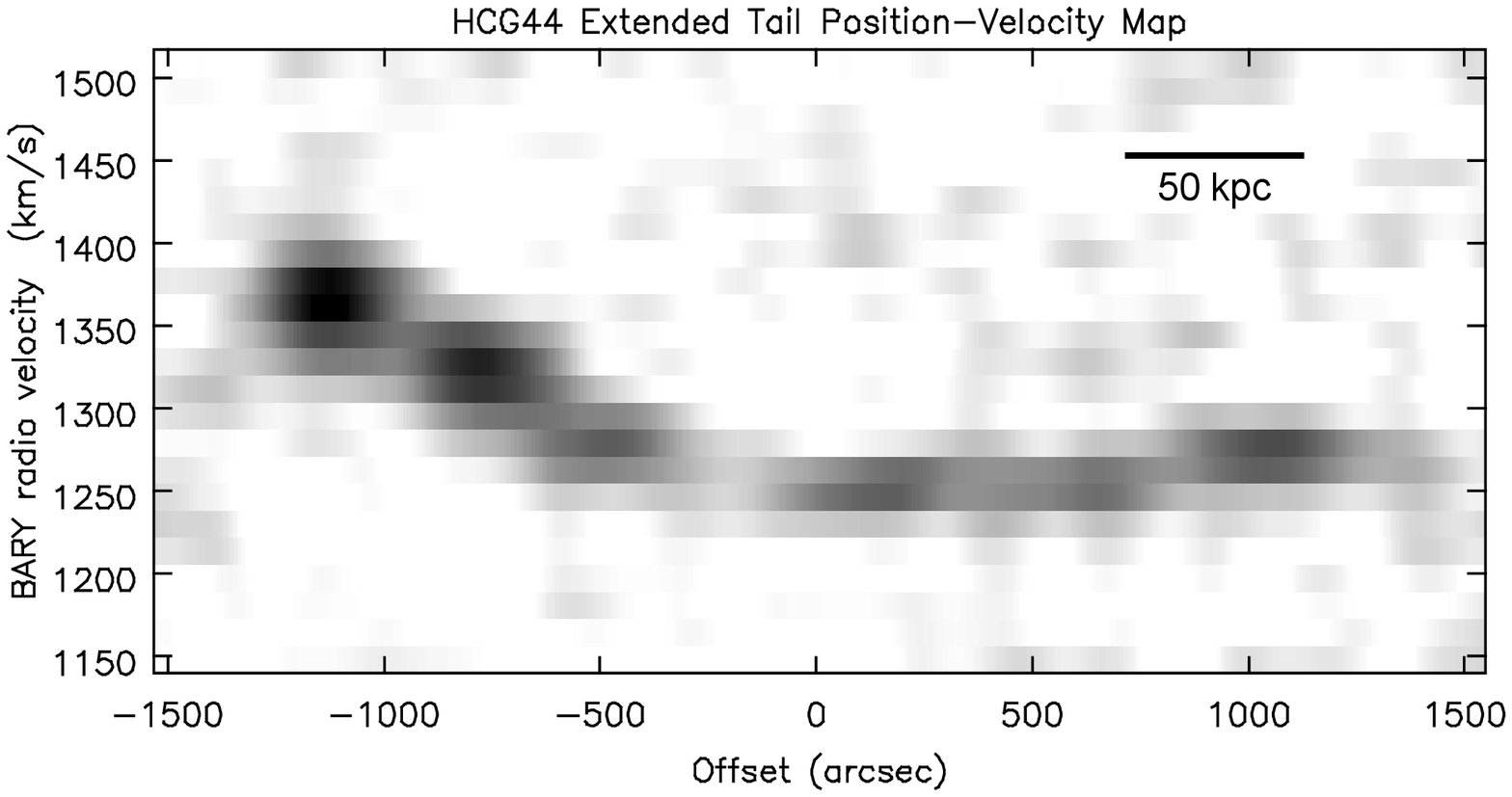}
\caption{A position-velocity slice taken along the main body of the \HI\ tail.  The data is summed over a slice 5.2 arcmins wide that cover the major clouds within the tail.  The velocity spans $\sim120$--150~\kms, and the dispersion is of order $25-55$~\kms across the full 350 kpc length of the tail.}
\label{pvmap}
\includegraphics[width=\columnwidth,clip,trim=00 300 0 150]{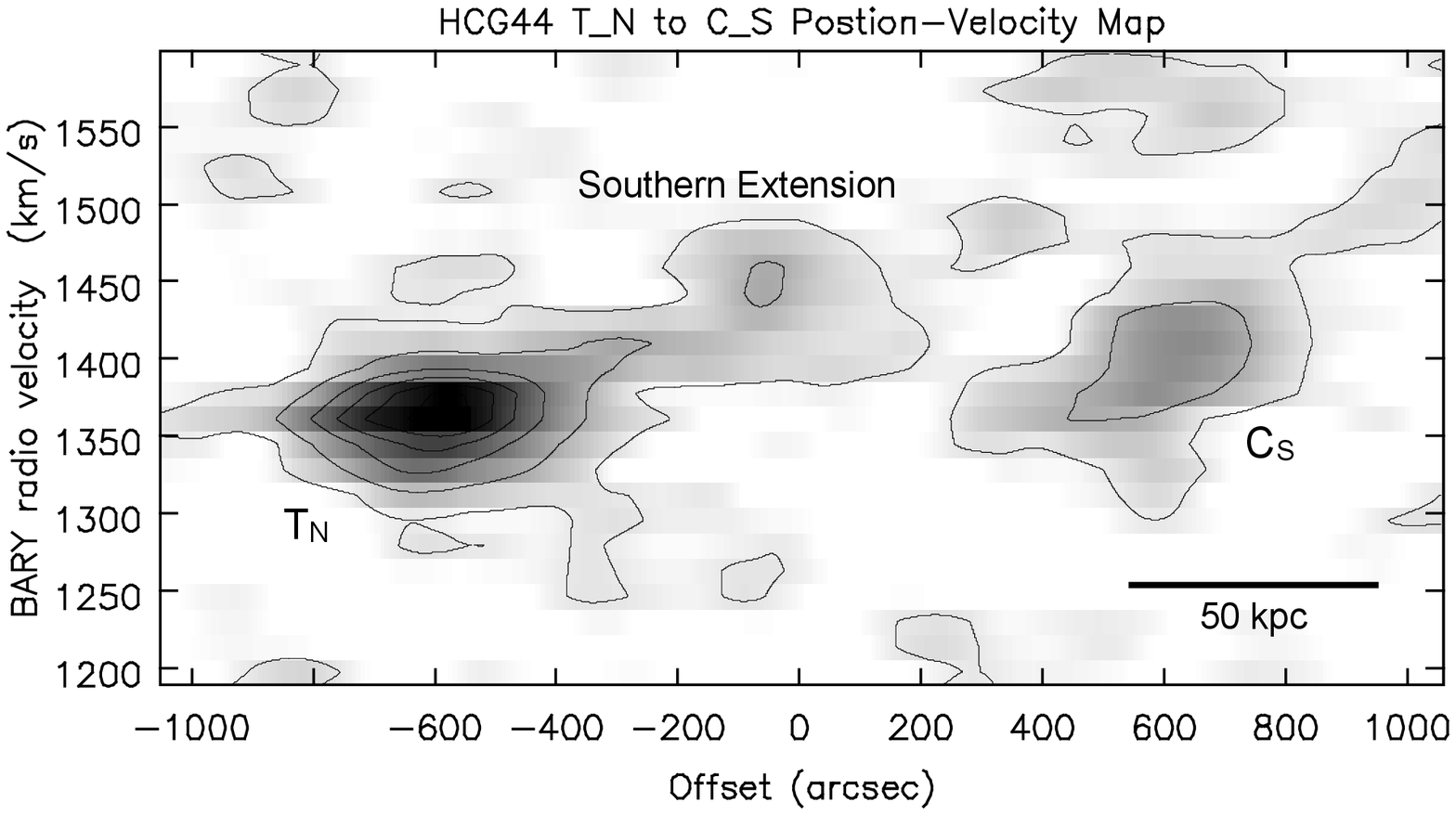}
\caption{A position-velocity slice taken between $T_{\text{N}}$ and $C_{\text{S}}$.  The data is summed over a slice 1.2 arcmins wide.  The contours have arbitrary values and are included to guide the eye.  There is no strong evidence for an HI bridge to the compact group beyond the southern extension of $T_{\text{N}}$.}
\label{pvmap2}
\end{figure}

Figure \ref{tailvel} shows the intensity weighted velocity map of the extended \HI\ tail.  The extended tail is contiguous in velocity with a shallow gradient that is parallel with the long axis of the tail from $\sim1365-1245$~\kms.  In fact, the westernmost extent, at 1270~\kms, is not as blueshifted as the center of the tail at 1250~\kms. Meanwhile the most massive clumps in the tail, which increase in mass to the east, are continuous from $1260-1365$~\kms.  Figure \ref{pvmap} shows the extended tail forms a shallow arc in velocity.  Despite the 350 kpc extent of tail, it is quite narrow in velocity with a dispersion of only $\sim25-55$~\kms.  In the most massive clumps, $T_{\text{N}}$ has a dispersion of roughly $45-55$~\kms, which peaks in cloud 3.  Beyond $T_{\text{N}}$, clouds 6, 7, and 8 have a dispersion of $25-35$~\kms.

\subsubsection{Tail origins?} \label{origins}

While \HI\ is invaluable for recognizing a history of past interactions, disentangling the chain of events is more complicated.  The origin of the extended \HI\ tail associated with HCG~44 is no less mysterious.  \citet{Ser13} ruled out the likelihood that the tail is the result of ram pressure from the lack of diffuse X-ray emission. The authors discussed two possible scenarios.  First, the gas was pulled out from NGC~3162 in a close passage with the compact group. The elongated morphology of $T_{\text{N}}$ and the HIPASS detection suggest a tenuous connection between the tail and NGC~3162, which lies approximately 620 kpc northwest of HCG~44 and the same distance from cloud 2 of the tail.  Second, the gas originated from one of the galaxies currently within HCG~44 that has interacted with the group gravitational potential.  Within HCG~44 both the stellar and resolved \HI\ component of NGC~3187 show a strong S-shaped distortion (Figure \ref{wise} and the WSRT contours in Figure \ref{mom0}) suggesting a strong tidal interaction.  Meanwhile, in SDSS optical images NGC~3190 has a strong dust lane with a swept-back morphology looking as though it has passed from the direction of $T_{\text{N}}$, and the galaxy has strong infrared H$_2$ emission which has been attributed to shocked gas \citep{Clu13}.

An outstanding question has been whether \citet{Ser13} catch a glimpse of the high column density gas in an \HI\ bridge which connects $T_{\text{N}}$ to the compact group.  If such a bridge exists, it should be visible in both our KAT-7 and combined images in the most sensitive region of the cubes.  A connection would add weight to the second origin scenario, and we could trace from which HCG~44 member galaxy the gas was stripped.  We went through a concerted effort to uncover such a bridge, going through multiple iterations of targeted cleaning of the KAT-7 cube during the imaging process, inspecting cubes at all stages of post-processing, including the KAT-7 cube, the final data cube described in Section \ref{combo}, the spatially smoothed cube used for masking, and previous versions of the cubes at higher velocity resolution. Finally we inspected position-velocity slices of the cubes at various angles from NGC~3190 and NGC~3187 towards the high column density clumps of $T_{\text{N}}$.  Ultimately, we confirm faint continuous emission between cloud 1 and cloud 2 of $T_{\text{N}}$, which makes up the southern extension, but we found no convincing evidence for an \HI\ bridge connecting $T_{\text{N}}$ to any of the HCG~44 members at our detection limits (e.g.~Figure \ref{pvmap2}).

The gap in \HI\ emission between $T_{\text{N}}$ and HCG~44 may be explained if the H$_2$ emission in NGC~3190 is not the result of shocked gas due to collisions with tidal debris, but comes from the presence of an active galactic nucleus (AGN; \citealt{Mar10}) due to jet-interstellar medium (ISM) interactions \citep{Og10}.  The orientation of NGC~3190 is such that ionizing radiation emanating from the nucleus would have a clear path in the direction of where a bridge between HCG~44 and $T_{\text{N}}$ would exist.  Thus, it is possible that we do not detect an \HI\ bridge because it has been blown out or ionized, similar to gaps seen in \HI\ tidal features such as around NGC~520 \citep{Hib00}.

On the other hand, images from ALFALFA which cover a larger area of sky show increasing evidence for a connection between the extended tail and NGC~3162 \citep{Lei16}.  It is not obvious that the gas must have originated from NGC~3162 because there is still a gap in the observed \HI\ distribution between the tail and NGC~3162.  However, we note that the morphology of the extended tail is consistent with that of gas pulled out of galaxies in other high velocity encounters ($v\sim1000$~kms), such as VIRGOHI~21 near NGC~4254 \citep{Duc08}.  In these simulations, the \HI\ tidal tails that form are long, narrow, and have a low stellar content.

There are further considerations, although it is unclear as to how they should be interpreted.  First, the tail is resolved into clumps and does not have a large velocity gradient as is seen in the \citet{Duc08} simulations, although the latter may be projection effects.  Second, the most massive \HI\ clouds are those furthest from NGC~3162, and closest to HCG~44.  The most massive clouds to result from a tidal interaction are likely those pulled out near periastron, where tidal forces are the greatest.  Third, the tail is the messiest to the southeast, close to HCG~44.  This could be because the gas originated within the compact group, or if the tidal tail was pulled out from NGC~3162 it is now interacting with the compact group.  Finally, we note that NGC~3162 does not suffer from \HI\ deficiency \citep{Lei16}, although the uncertainty in expected \HI\ mass for a galaxy of a given morphology and diameter is enough that we cannot rule out NGC~3162 as the donor.  

In any case, it is probable that multiple interactions have occurred in the system. A third scenario is that multiple interactions within the compact group loosened the cold gas of the galactic ISM and allowed it to form a long tail in a high speed encounter with NGC~3162, as modeled by \citet{Duc08}.  The internal interactions shake up the \HI\ gas which is then not bound as tightly in the shallow group potential, and would also account for the disturbed morphology of NGC~3190.  The overall \HI\ deficiency of galaxies in the compact group would be the result of their history of group membership and gas depletion through star formation and gas processing, as proposed in other group environments by \citet{VM01} or \citet{HW13}.  If the gas in the extended tail originated in an HCG~44 galaxy, it has not had to travel as far from its starting point as if it originated in NGC~3162.  In short, we present a number of possible scenarios for the origin of the extended tail, but the KWA data presented here is unable to say conclusively which is correct.

\subsubsection{Collapse or Dispersal?} \label{fate}

Despite the uncertainty in its origin, we investigate the possible fate of the \HI\ tail.  Our data shows that it has a clumpy distribution with 7-8 knots of emission.  If we consider each knot as a distinct \HI\ cloud, we note that the clouds with the highest peak column density and greatest velocity dispersion are the easternmost clumps of the tail. 

Tidal interactions are known to sometimes produce knots of star formation in stripped gas, and if these structures are self-gravitating, they are referred to as tidal dwarf galaxies \citep{BH92,DM99}. 
Assuming the clouds are spherical, and the observed velocity dispersion is the thermal velocity of the gas, we can test to see if they will collapse.  The easternmost knot is $2.2\times10^8$~\msun\ and has a velocity dispersion of $\sim50$~\kms.  The Jeans' length is $R=3/5\,GM/\sigma^2\approx0.2$ kpc.  The WSRT appears to resolve this cloud to the size of 5 arcminutes which corresponds to 36 kpc (Figure \ref{mom0}, or Figure 2 in \citealt{Ser13}) so it is unlikely that stars are forming out of gas in the tail.  
Additionally, despite deep images from CFHT/MegaCam with a limiting surface brightness sensitivity of $\sim$28.5 mag arcsec$^{-2}$ in g-band \citep{Ser13}, there is no known stellar component in the optical, and nothing evident in WISE infrared imaging (Figure \ref{wise}).   If stars were pulled out of the galaxy concurrently with the gas, they would be broadly dispersed and potentially undetectable \citep{Hib00}.  Searches for planetary nebula line emission as a tracer for intragroup stellar light, conclude an upper limit of only 5\% for the diffuse light fraction within HCG~44 itself \citep{Agu06}.

On the other hand, we may investigate cloud survival in the context of photoionization.  Observations have shown that both the outer disks of galaxies and tidal debris are protected from the ionizing ultraviolet (UV) background if the gas is above a few $\times10^{19}$~cm$^{-2}$ \citep{Cor89,vanGor91,Hib00}.  Radiative transfer modeling suggests that clouds can survive in the intergalactic medium if they have a column density of at least $2\times10^{19}$ cm$^{-2}$ \citep{Mal93,Bor15}.  In our total intensity map (Figure \ref{mom0}) the extended envelope of gas in the \HI\ tail is closer to $1\times10^{19}$ cm$^{-2}$ or less, but this may be an underestimate since the clouds themselves are largely unresolved. The peak column density of clouds 2 and 3 are $2.3\times10^{19}$ and $1.8\times10^{19}$ cm$^{-2}$, respectively.

\citet{Bor15} point out that a 1 kpc cloud with an ISM density of 1 atom~cm$^{-3}$ expanding at 20~\kms\ would survive for $\sim$500 Myr before dropping to a density susceptible to photoionization by the UV background.  However, if the density of the IGM is of order a few to tens $\times10^{-4}$ cm$^{-3}$ the cold \HI\ clouds will reach pressure equilibrium before reaching the critical column density, and may survive longer.  These IGM densities have been inferred indirectly through the morphology of bent jets in galaxy groups \citep{Free11}.  

We find that both the column density and velocity dispersion of the \HI\ clouds decrease as one moves along the tail from east to west (see Figure \ref{pvmap} and Section \ref{hitail}).  If we consider that the gas has been pulled out of NGC~3187, as in \citet{Ser13}'s 3D toy model, and/or NGC~3190, the clouds may represent snapshots of evolution at different times after they have been pulled out of the galaxy.  In this case the \HI\ clouds become fainter as they disperse and the velocity dispersion decreases as what remains is the cold core of the cloud.  Our extrapolation of the proposed trajectory from the 3D model implies that the easternmost \HI\ clouds were stripped of order 500, 600, and 700 Myr ago.  Further, the clouds in the extended tail fall at intervals which would imply they were stripped approximately 850, 1000, and 1100 Myr in the past.  These lifetimes suggest that intragroup \HI\ clouds may have survived by being contained in pressure equilibrium with a warm intragroup medium that has not been directly detected, and which may extend out to quite large distances beyond the compact group--perhaps as part of the larger GH58 or NGC~3190 loose group to which HCG~44 apparently belongs \citep{GH83,vDr01,Mak11}.

Alternatively, we can make a simple estimate of the required survival time of the clouds if they are a result of a close encounter with NGC~3162.  NGC~3162 lies at a projected distance of 620 kpc from HCG~44, and at relative fly-by velocities of $600-200$\kms, the clouds would have had to survive between 1-3 Gyr.  The lower velocity limit seems challenging, given the long required lifetime of the clouds in the IGM.  Meanwhile, the high velocity limit has been shown to produce long tidal tails in cluster systems, but may be unusually high for a galaxy group system with few members and an otherwise narrow velocity dispersion (115-220~\kms\ depending on what is considered part of the group; \citealt{Lei16}, \citealt{GH83}).

Simulations of small \HI\ clouds survival in a Milky Way-like hot ($T=1-3\times10^6$ K; $n_H\approx1-3\times10^{-4}$ cm$^{-3}$) halo suggest that neutral \HI\ which has been stripped can last of order 80--200 Myr before it is disrupted or ionized \citep{Fer12,Heit09}.  The \HI\ clouds of these simulations have a range of mass from ($\sim10^{4-8}$~\msun), but \citet{Put12} summarizes that total mass seems to be one of the largest factors which increase their lifetime.  The hydrodynamic simulations also include the effects of ram pressure stripping, in addition to photoionization.  If the effects of ram pressure are less severe on the HCG~44 intragroup clouds \citep{Hib00}, perhaps they are able to live longer.  If they have already lived in the IGM for 1 Gyr, they probably started with a larger initial mass than what we currently observe.

\subsection{\HI\ in Group Evolution}

\HI\ is a key tracer of physical mechanisms at work in galaxy groups as the gas is both the most extended component of pristine galaxies, and also most easily removed through gravitational and hydrodynamical interactions.  It can provide rigorous constraints on galaxy-galaxy interactions, group dynamics, and indirect evidence of the warm-hot IGM which remains otherwise undetected.  Tidally stripped \HI\ in groups appears to survive as both filamentary structures and a diffuse component \citep{VM01,Bor10}. This gas can provide a reservoir for re-accretion through so-called ``rejuvenation events", but collisions with this intragroup medium may also produce shocks in the interstellar medium of galaxies \citep{Clu10,Clu13}. Associated shock heating may lead to further stripping, or the formation of molecular hydrogen through gas compression at the leading edge of the shock\citep{App06,Clu13}.  To date the preponderance and evolutionary effects of these processes remain relatively unexplored and poorly understood.  In HCG~44, if previously detected \Ht\ is the result of shocks due to collisions with the IGM, the intragroup medium doing the work remains undetected in our deep \HI\ data.  On the other hand, if the \Ht\ emission is due to jet-ISM interaction shocks from an AGN in NGC~3190, it may explain the absence of an \HI\ bridge between HCG~44 and the extended \HI\ tail.

Qualitative analytical arguments for \HI\ cloud survival have estimated cold cloud lifetimes of 500 Myr once gas has been removed from a galaxy disk \cite{Bor15}.  However, if the orbit of NGC~3187 modeled by \citet{Ser13} is representative of a group member interacting with the HCG~44 group potential, the \HI\ clouds at the end of the extended tail may have already survived for up to 1 Gyr.  If the gas originated in NGC~3162, the clouds have lived in the circumgroup medium for 1-3 Gyr.  This result warrants further simulation to understand how the balance of physical processes between the cloud and the circumgroup medium allows for cloud survival.  Unfortunately, the warm-hot intragroup medium has thus far evaded direct detection so the density and temperature of the intragroup medium is constrained more by what we do not see than what we do (e.g.~\citealt{Dave01,Free11}).  However, our result demonstrates the potential of \HI\ to provide even stronger constraints on the modeling of galaxy group properties.

Future wide-area \HI\ surveys (e.g. WALLABY, DINGO, MIGHTEE-HI, and APERTIF surveys) will uncover many more nearby \HI\ rich groups for deep follow-up.  
A number of preliminary results from deep KAT-7 observations (Oosterloo et al, in prep; Lucero, private communication) and from ASKAP \citep{Ser15}, show that extended and low-surface brightness \HI\ features may be even more common in galaxy groups than previously realized.  Tidal interactions produce material that lead to viscous stripping of the interstellar medium from galaxies, and this process may be important in loose groups in addition to compact groups.

Finally, with the expected increase in the gas fraction of galaxies as one moves to higher redshift, group interactions may play an even more significant role in the evolution of galaxies in the past than what is seen locally.  In other words, the effect of environment when galaxies were typically gas-rich ($z\sim1-2$) may be a key consideration at a time when the universe was experiencing elevated star formation and more efficiently building stellar mass compared to local systems.

\section{Conclusions} \label{conclusions}

We observed the region around HCG~44 and the giant \HI\ tail discovered by \citet{Ser13}, which extends to the north and west of the compact group, to investigate the extent of intragroup \HI.  We combined a deep two-pointing mosaic from KAT-7 with imaging from the ALFALFA survey, and a single pointing observation of WSRT.  Despite the very different nature of the telescopes, we successfully combined the data by converting the image units of Jy beam$^{-1}$ to column density per unit velocity, atoms cm$^{-2}$(\kms)$^{-1}$, weighting by the sensitivity of the respective observations, and adding them in the image plane.

We find no evidence for copious amounts of neutral \HI\ in the IGM within the core of the compact group, which suggests the \Ht\ emission observed by \citet{Clu13} in NGC~3190 may be the result of AGN activity and jet-ISM interaction rather than shocked gas from NGC~3190 plunging through the IGM.  The presence of an AGN, or star formation in the nucleus of NGC~3190 could account for the lack of an \HI\ bridge between HCG~44 and the \HI\ tail, if HCG~44 is where the extended \HI\ tail originated from.

The combined \HI\ data show that the \HI\ tail extends out to at least 450 kpc and is resolved into 7--8 clouds at the resolution of KAT-7.  We find more \HI\ mass in all parts of the tail as compared to previous WSRT observations, such that the total \HI\ mass in the tail is $1.1\times10^9$~\msun.  However, the gas in the tail is still not able to account for the overall \HI\ depletion in the compact group.  The remaining gas is either too diffuse to be measured with current facilities, or more likely has been ionized and may exist in a warm-hot group halo that is also undetected in X-rays.

We find that if the gas in the tail is stripped from one of the HCG~44 member galaxies, and the timescale for stripping proposed in \citet{Ser13} is correct, the gas must have been able to live in the intragroup medium without being ionized for 0.5-1 Gyr, which is longer than previously predicted.  If the gas originated in NGC~3162, it must have been produced in a high speed encounter, and even then must have survived in the IGM for well more than 1 Gyr.  Regardless of the origins of the \HI\ tail, our results demonstrate indirect evidence for a surrounding medium which has confined the expansion of the \HI\ clouds and prevented their destruction by background UV radiation.  Targeted simulations are required to explain how these \HI\ clouds can survive for long periods of time in the IGM.

\section*{Acknowledgements}

The research of KMH has been supported by the European Research Council under the European Union's Seventh Framework Programme (FP/2007-2013)/ERC Grant Agreement nr.~291531.
MEC acknowledges support from the South African National Research Foundation.
The work of SY was supported by the 2015 ASTRON/JIVE Summer Student Programme.  
The work of CC is based upon research supported by the South African Research Chairs Initiative (SARChI) of the Department of Science and Technology (DST),  the Square Kilometre Array South Africa (SKA SA) and the National Research Foundation (NRF).
We thank Thijs van der Hulst, DJ Pisano, Lourdes Verdes-Montenegro for useful discussions.

This work is based on observations obtained at the KAT-7 array which is operated by the SKA SA on behalf of the National Research Foundation of South Africa.

The authors would like to acknowledge the work of the entire ALFALFA collaboration team in observing, flagging, and generating the data products used in this work.  The ALFALFA team at Cornell is supported by NSF grants AST-0607007 and AST-1107390 to RG and MPH and by grants from the Brinson Foundation.

This paper made use of observations obtained with the Westerbork Synthesis Radio Telescope, which is operated by the ASTRON (Netherlands Foundation for Research in Astronomy) with support from the Netherlands Foundation for Scientific Research (NWO).

This publication makes use of data products from the \textit{Wide-field Infrared Survey Explorer}, which is a joint project of the University of California, Los Angeles, and the Jet Propulsion Laboratory/California Institute of Technology, funded by the National Aeronautics and Space Administration (NASA).












\bsp	
\label{lastpage}
\end{document}